\documentclass[aps,preprint]{revtex4}
\usepackage{graphicx}
\def\BibTeX{{\rm B\kern-.05em{\sc i\kern-.025em b}\kern-.T
    08em\kern-.1667em\lower.7ex\hbox{E}\kern-.125emX}}

\begin{document}
\title{Self-fields in thin superconducting tapes: implications to the thickness effect in coated conductors}

\author{Alvaro Sanchez$^1$, Carles Navau$^1$, Nuria Del-Valle$^1$, and Du-Xing Chen$^{2,1}$}
\affiliation{$^1$Grup d'Electromagnetisme, Departament de F\'{\i}sica, Universitat Aut\`onoma de Barcelona, 08193 Bellaterra,
Barcelona, Catalonia, Spain\\
$^2$Instituci\'{o} Catalana de Recerca i Estudis Avan\c cats (ICREA), Passeig Llu\'{\i}s Companys 23, 08010 Barcelona, Catalonia, Spain}

\maketitle

{\bf Most applications of superconductors, such as power transmission lines, motors, generators, and transformers, require long cables through which large currents circulate. Impressive progress has recently been achieved in the current-carrying capability in conductors based on high-temperature superconductors. Coated conductors are likely the best examples, consisting of very good quality thin layers of YBa$_2$Cu$_3$0$_{7-\delta}$ (YBCO) superconductor grown on top of a metallic tape with some intermediate layers. However, there is an important problem for achieving large currents: a large decrease in transport critical-current density $J_{\rm c}$ when increasing film thickness has been observed in coated conductors made by all available techniques  \cite{foltynnature,larbalestiernature,luborsky,foltynAPL93,foltynAPL99,kang,emergo,civale,foltynAPL05,kim,takahashi,foltynsust2009}. Here, we theoretically explain the nature and the ubiquitous presence of this so-called thickness effect by analyzing the self-field created by the transport currents in the superconductor, assuming a realistic field-dependent $J_c$. This knowledge can help finding new ways to improve transport current in thick superconducting films.  }

The most typical manifestation of the thickness effect occurs when measuring the transport $J_{\rm c}$ of superconducting tapes at self-field (no externally applied magnetic field) as function of the tape thickness. 
Several explanations have been proposed to partially account for some of the effects, such as enhancement of flux pinning near the heteroepitaxial interface between YBCO and the substrate or interlayer \cite{foltynsust2009}, difference in intergrain coupling for films with different thicknesses \cite{solovyov,feldmann}, or non-uniform distribution of pinning sites across the thickness \cite{feldmannapl} . However, the fact that this effect is common for all combination of substrates, preparation technologies, and HTS deposition techniques \cite{vmpan,peer} suggests  
that there is an underlaying phenomenon that is present in all the studied cases. In this work, we argue that the effect of the self-field of transport currents in thin films is an underlying basis for all the thickness dependence experiments in coated conductors. The existence of this 'universal' geometry-related thickness effect does not mean that the other mentioned effects (inhomogeneities, interface effects, etc) do not play an important role in particular cases. However, only after understanding the geometrical effects can the rest of phenomena be properly studied.

We consider an infinitely long type-II superconducting strip of rectangular cross-section located at $-w/2\leq y\leq w/2$ and $-t/2\leq z \leq t/2$, which obeys the critical-state model, with a field-dependent critical-current density $J_{\rm c}(H_{\rm i})$, where $H_{\rm i}$ is the modulus of the total magnetic field. 
We calculate the self-field and the local current density at every point of the superconductor cross-section by an iterative process described in the Methods section (similar procedures were used in \cite{bbclem,rostila}).

What can the model explain about the thickness effect? To answer this question we plot in Fig. 1 as dots the experimental points corresponding to collected data for different YBCO coated conductors, obtained from the review paper Ref. \cite{foltynnature}. These data can be considered a typical systematic study of thickness dependence in state-of-the-art YBCO conductors. The line in Fig. 1 corresponds to the $J_{\rm c}$ calculated by our model, assuming an intrinsic Kim-type $J_{\rm c}(H_{\rm i})$ dependence 
\begin{equation}
\label{eq.kim}
J_{\rm c}(H_{\rm i})= {{J_{\rm c0}}\over{1+{H_{\rm i}}/{H_0}},}
\end{equation}
with values $J_{\rm c0}=12$ MA/cm$^2$ and $\mu_0 H_0=0.006$ T, and using the actual film dimensions of the experiments (Kim model is chosen becauese it has been shown to fit well the $J_{\rm c}(H_{\rm a})$ measured in thin YBCO films \cite{suenaga}). The calculated critical-current density  follows well the drop observed in the thickness-dependence experiments. The standard thickness dependence is therefore very well reproduced just by assuming a (realistic, see below) $J_{\rm c}(H_{\rm i})$  function and calculating the transport current at self-field for the exact dimensions used in the experiments, without any extra parameter. The explanation has no need resorting to a spatially dependent $J_{\rm c}({\bf r})$ function including an extra  distance parameter, as in some previous models \cite{foltynAPL05,foltynsust2009}; although this spatial dependence may indeed by present in some cases, it is not logical that it appears for so many sample preparation conditions in which the effect has been observed \cite{peer}.

The next question to answer is how the Kim-model $J_{\rm c}(H_{\rm i})$ function we have used represents the actual field dependence in the real samples.  In Fig. 2 we plot as  dashed line the $J_{\rm c}(H_{\rm i})$ dependence used in Fig. 1 [Eq. (1)] and as solid line the expected measured $J_{\rm c}$ as a function of applied field $(H_{\rm a})$ calculated by our model taking into account the self-field, for the case of thickness $t=1\;\mu$m. When comparing the calculated $J_{\rm c}(H_{\rm a})$ 
with a typical experimental dependence (as in Fig. 8 of the same review paper \cite{foltynnature}), one can see that the theoretical curve has the same behavior as the experimental one, showing a plateau region up to an applied field $\mu_0H_{\rm a}\sim 0.01 $  T and then decreasing with applied field \cite{foltynnature}.
In contrast, the intrinsic $J_{\rm c}(H_{\rm i})$ dependence is larger at low fields than the plateau value and merges with the calculated $J_{\rm c}(H_{\rm a})$ for fields larger than around 0.03 T, because in this region the self-field created by currents is much less than the applied field value, so the internal field is basically $H_{\rm a}$ \cite{chengoldfarb,peak,bbclem,rostila}.

Another conclusion can be obtained from the results in Fig. 2: although the measured value of $J_c$ at applied field $H_{\rm a}$=0 is the plateau value $J_{\rm {c,plat}}\sim$3.3 MA/cm$^2$ for a thickness $t=1\mu$m, our model shows that intrinsic $J_{\rm c}(H_{\rm i})$ can have larger values (a maximum of 12 MA/cm$^2$ in this case). This means that, following our model, when measuring transport current at self-field in the coated conductor, there are regions that actually have a current density larger than the value of the plateau. This is shown in Fig. 3 where we plot the local values of $J_{\rm c}$ (Fig. 3a) and $\mu_0H_{\rm z}$ (Fig. 3b) for the conductor with $t=1\mu$m in the conductor midplane, corresponding to Fig. 2, for different $H_{\rm a}$ values. One can see that in the central region of the sample the local $J_{\rm c}$ exceeds the plateau value $J_{{\rm c,plat}}\sim$3.3 MA/cm$^2$ (dotted line); actually, exactly at the conductor center, the magnetic field is zero (according to Ampere's law) and therefore current density has there the maximum value of the intrinsic $J_{\rm c}(H_{\rm i})$, 12 MA/cm$^2$. The fact that there are regions in the sample with local $J_{\rm c}$ larger than the plateau value opens a path for improving transport $J_{\rm c}$ in these conductors.
It is also interesting to note that the maximum magnetic field in the sample for $H_{\rm a}$=0, which occurs at film edges, corresponds rather well to the product $J_{{\rm c,plat}} t$ \cite{bbclem}, and not to $J_{{\rm c0}} t$ (in the studied case $J_{{\rm c0}}=12$ MA/cm$^2$, around four times larger than $J_{{\rm c,plat}}\sim$ 3.3MA/cm$^2$) . When the applied field is increased, the maximum in local $J_{\rm c}$ first moves to one side and then eventually smears out and local $J_{\rm c}$ and field become both basically constant.

The model for the thickness effect presented in this work also explains the results deduced from the incremental model presented in \cite{foltynAPL05,foltynsust2009}. This model artificially assumed a spatially dependent $J_{\rm c}({\bf r})$ function, which decreases linearly with the vertical distance to the substrate up to a given distance, and then it is constant. From our results, the drop of incremental $J_{\rm c}$ naturally arises from the field dependence of $J_{\rm c}$, with the property that $J_{\rm c}$ would not decrease in general with the distance from the substrate but with the distance from the center of the sample. 

Because the described effects are geometrical and then always present in films, is it therefore impossible to have a thickness-independent $J_{\rm c}$? The answer is not; a possible solution is to try to achieve superconductors such that its intrinsic $J_{\rm c}(H_{\rm i})$ is roughly constant from 0 to at least the self-field, of the order $J_{{\rm c,plat}} t$. Probably this is the situation in some measurements of films that show almost no thickness dependence of $J_{\rm c}$ \cite{foltynsust2009}.

In view of the above results, we can conclude that an explanation for the ubiquitous thickness effect has been given. It is because of its geometrical nature that it is present in striking similar form for a variety of deposition techniques (PLD, MOCVD, {\sl ex situ} MOD films, etc. \cite{peer}). 
However, we would like to remark that our model, although it describes the typical thickness dependence experiments, does not explain some particular experimental data. These cases may thus become the key for finding new strategies of optimizing current capabilities of thick tapes. In one of such cases \cite{foltynAPL05}, Foltyn et al present large values of $J_{\rm c}$ for multilayers composed of alternating YBCO and CeO$_2$ layers. The values of $J_{\rm c}$ for a (thick) multilayer composed of six 0.55$\mu$m-thick layers of YBCO, with a layer separation of only around 40nm, were very close to the values of a single (thin) YBCO layer. This cannot be explained by our model, since the six YBCO layers separated such short distances should have a $J_{\rm c}$ similar to that of the whole block, and smaller than that of a single layer. Another example of results which our model cannot explain is the observation of thickness dependent $J_{\rm c}$ at fields of the order of 1T or more, far above the self-field of currents \cite{kim}. Since our model for homogeneous samples fail there, this probably indicates that in these cases spatial inhomogeneities are indeed present, such as enhancement of $J_{\rm c}$ at the interface \cite{foltynAPL05}. Also, the presence of other effects such as anisotropy or surface barriers may yield results departing for our model predictions, therefore signaling the presence of these phenomena. The described examples show that there are very promising routes to increase $J_{\rm c}$ in thick YBCO samples. However, these strategies can only succeed if the geometrical effects, for which we have provided analyzing tools in this work, are properly taken into account.

In conclusion, we have explained the existence of a thickness effect in many measurements in superconducting tapes and why this effect appears in samples prepared and measured under many different conditions. Only after this geometrical effect is correctly analyzed in a given experiment, the intrinsic properties of the studied superconductor can be extracted, and in general paths towards large values of transport current in thick superconductors can be properly explored.

We thank Anna Palau, Carlos Monton, Teresa Puig, Xavier Obradors, Alejandro Silhanek, Joffre Gutierrez, and Boris Maiorov for conversations. We thank Consolider Project NANOSELECT (CSD2007-00041) for financial support.

%\end{references}

\null
\newpage

%FIGURE 1
\begin{figure}[htbp]
	\centering
		\includegraphics[width=1.00\textwidth]{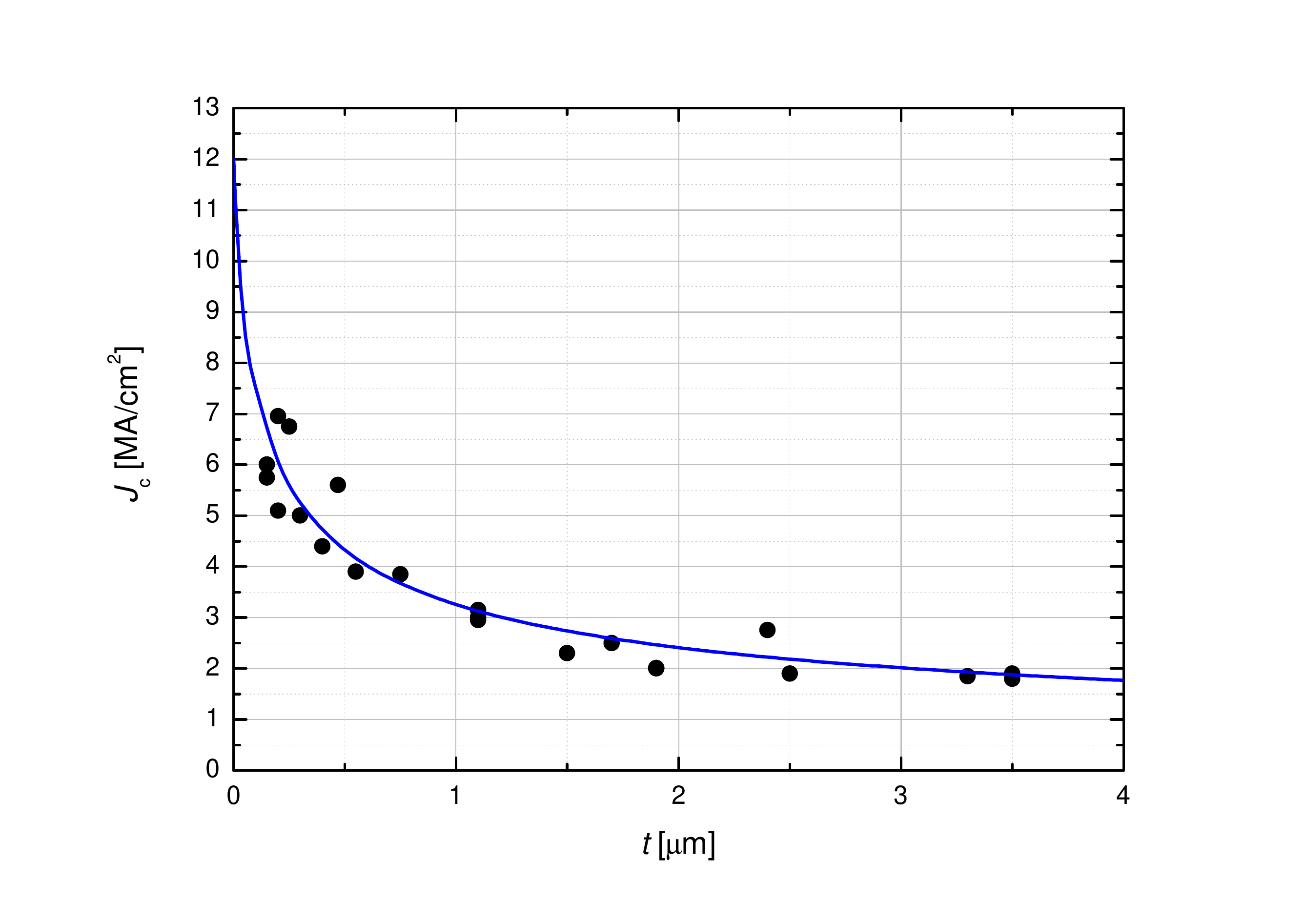}
	\caption{Transport critical-current density $J_c$ of coated-conductor YBCO films of width 200$\mu$m as function of film thickness $t$. Points are experimental data obtained from Ref. [1] and solid line is the theoretical calculation.}
	\label{fig.1}
\end{figure}

\begin{figure}[htbp]
	\centering
		\includegraphics[width=1.00\textwidth]{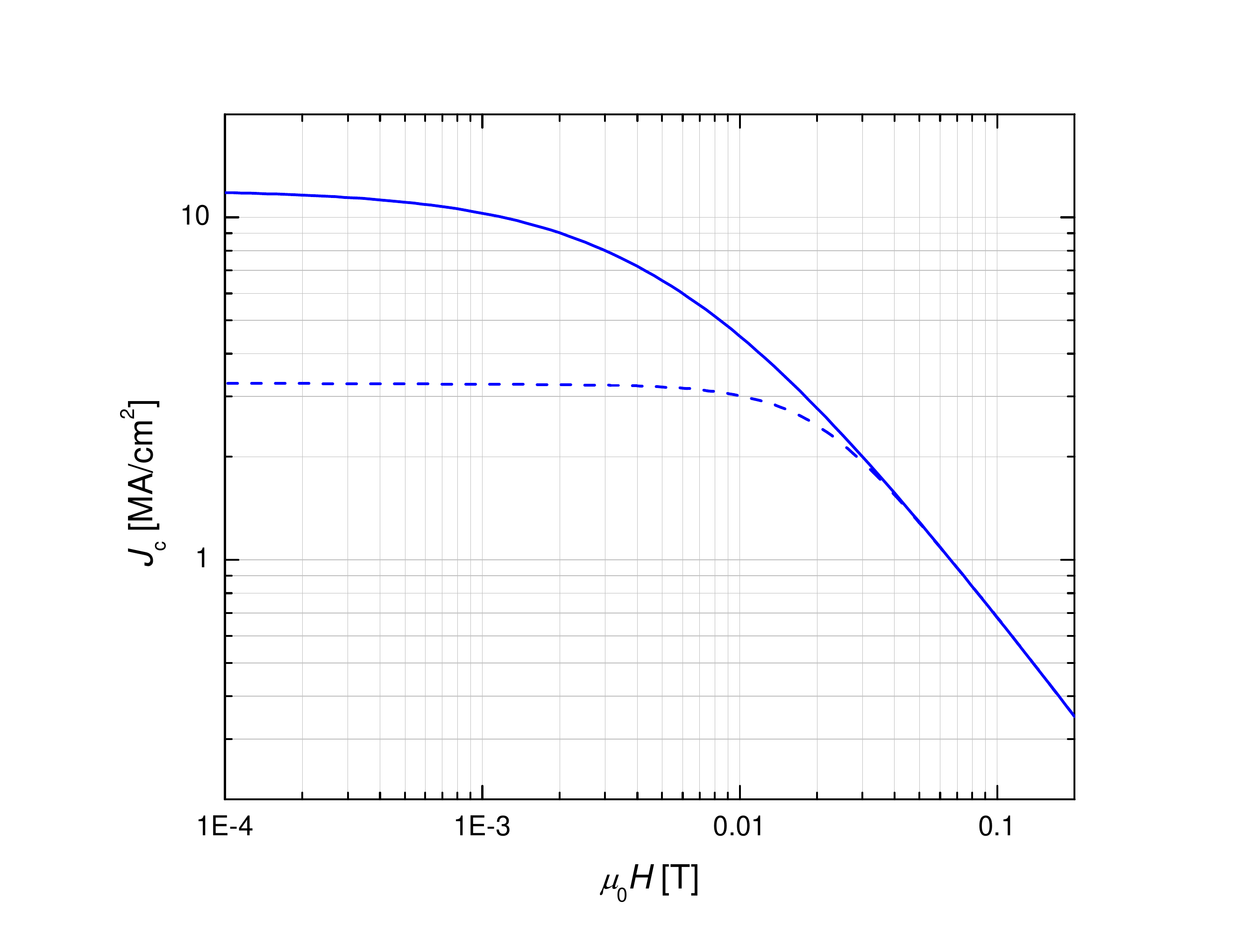}
	\caption{Critical-current density $J_c$ of coated-conductor YBCO films of width 200$\mu$m and thickness 1$\mu$m as function of applied field $\mu_0H_{\rm a}$. The solid line is the theoretically calculated $J_c$ and the dashed to the assumed $J_c$ dependence as function of internal $\mu_0 H_{\rm i}$ field [Eq. (1)].}
	\label{fig.2}
\end{figure}

\begin{figure}[htbp]
	\centering
		\includegraphics[width=1.00\textwidth]{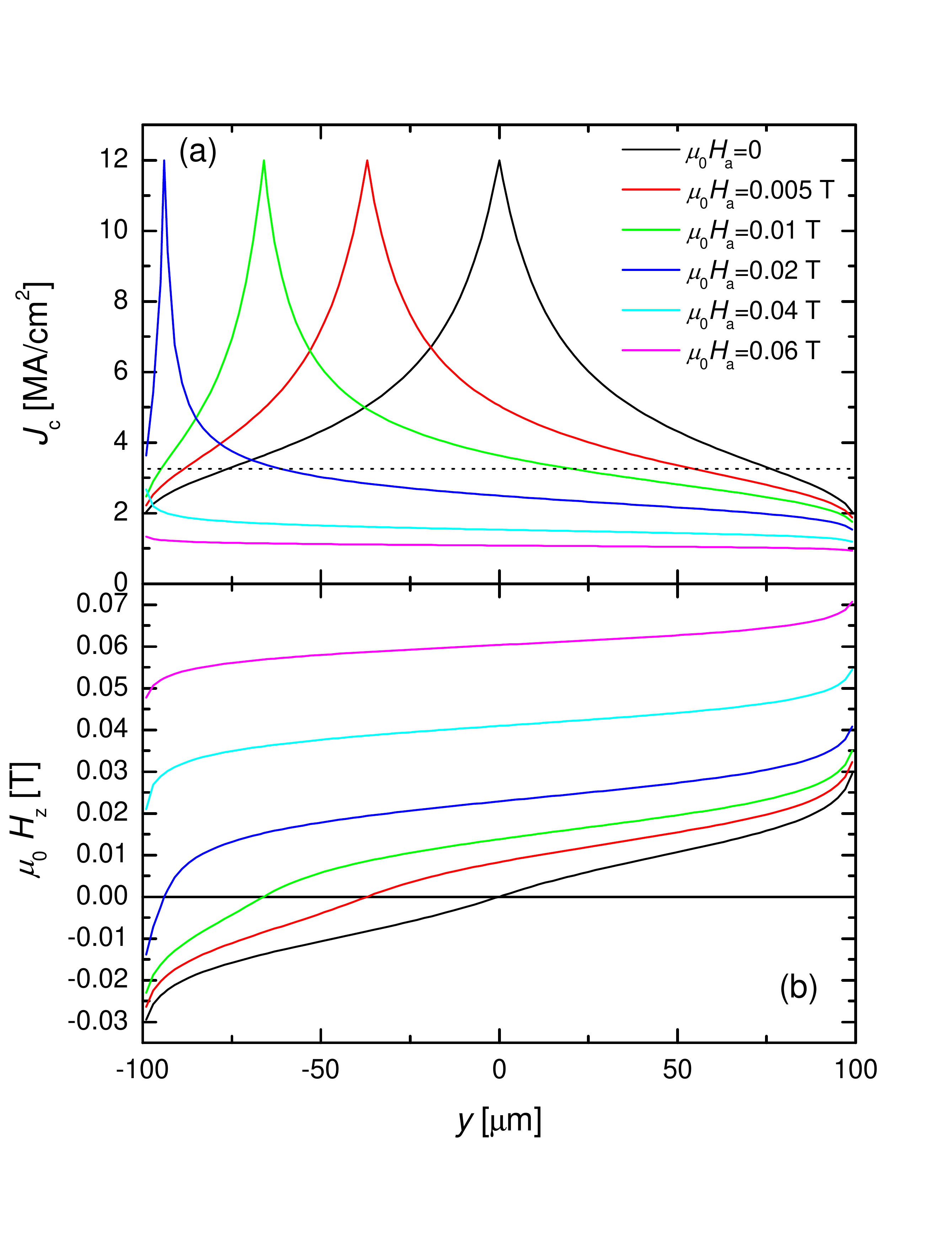}
	\caption{Calculated local critical-current density $J_c$ (a) and vertical component of magnetic field $\mu_0 H_z$ (b) as function of horizontal position for a coated-conductor YBCO film of width 200$\mu$m and thickness 1$\mu$m (as in Fig. 2) in the film midplane, for appplied field values $\mu_0 H_{\rm a}$=0, 0.005, 0.01, 0.02, 0.04, and 0.06T. The thin dotted line in (a) shows the $J_c$ plateau value for this film.}
	\label{fig.3}
\end{figure}

\end{document}